# Dielectric Measurement of Powdery Materials using a Coaxial Transmission Line

Robert Tempke[1,2,4], Christina Wildfire[3,4], Dushyant Shekhawat[4], Terence Musho[1,2,4,*]

[1] Mechanical and Aerospace Engineering Department, Morgantown, WV 26506, USA.
[2] Oak Ridge Institute of Science and Education, Oak Ridge, TN 37830, USA.
[3] Leidos Research Support Team (LRST), 3610 Collins Ferry Rd, Morgantown, WV, 26505, USA.
[4] National Energy Technology Laboratory, 3610 Collins Ferry Rd, Morgantown, WV, 26505, USA.
*tdmusho@mail.wvu.edu.

**Abstract:** The following study investigates the use of a coaxial transmission line for determining the properties of powdery dielectric materials (1-10GHz). Four powdery materials with dielectric constants ranging from 3.5 to 70 ($SiO_2$, $Al_2O_3$, $CeO_2$, and $TiO_2$) were experimentally investigated at varying volume loading fractions. Powder particles were mixed with a paraffin matrix and properties of the powder were analyzed using ten mixing equations to verify their accuracy. These powder-paraffin composites were also modeled at varying volume loadings for comparison with experimental data to gain a better understanding of the interactions between the different phases. The optimal volume loading fraction was determined to be 10% for all powders tested. A metric for selecting the most well-suited mixing equation was proposed that involved taking the ratio of the particle dielectric to that of the matrix. This study ultimately provides guidance for experimentally measuring the dielectric properties of unknown powdery materials that have application for new devices that utilized powder-based dielectric materials.

## I. INTRODUCTION

The experimental characterization of powdery materials is critical to new devices and process developments at GHz frequencies. While the physics of wave-matter interaction at microwave wavelengths has been well established, the experimental measurement technique to derive the dielectric properties of these powder materials is often subject to considerable random error. These random errors typically stem from the inconsistency of the sample preparation method and uncertainty in the general application of mixture equations used to derive the dielectric properties. In providing a solution, this study considers the most commonly used sample preparation methods and mixture equations to provide justification for minimizing uncertainty in determining the dielectric properties of powdery materials.

One of the targeted applications of these powdery materials is for electromagnetically driven chemical processes at microwave frequencies. This approach is being applied to a larger array of gas and solid-phase chemical processes in the aim of achieving new processing windows that require less energy. With new processing windows, comes new distribution of products that permit small, modular reactors to be realized. In designing these new processes, the use of catalytic materials that are tailored for both microwave absorption and chemical species selection are required. Often these materials are powder-based as they increase the specific density of reaction sites and decrease the dipole continuity of the material resulting in localized field enhancement. By influencing the dipole environment, the effective properties of the material are subsequently altered. It is hypothesized that a reactor design engineer could use this knowledge to their advantage to tailor the dielectric properties of a given catalyst powder for a given reaction process. In addition to controlling the dipole continuity, which is governed by the volume fraction of the powders used in a catalyst mixture, there are several other physical attributes that can be tailored to modulate the dielectric properties. These attributes include, particle shape (sharp edges), size (diameter), electrical conductivity, and bulk dielectric properties [2]. In order to understand how these physical changes in the powder correlate with changes in dielectrics it is necessary to evaluate existing methodologies and propose best practices to follow.

There are several experimental measurement techniques to measure the dielectric properties of powdery materials. The most common techniques being a rectangular free-space waveguide, an open-ended probe, or a coaxial transmission line [3]–[5]. The transmission line is the preferred method as it provides a precise measurement (low random error) for a large range of dielectric materials and frequencies. Both the free-space and transmission line methods permit a two-port transmission analysis of the material, which allows the complex form of both the permittivity and permeability to be determined. For this study, the focus will be on non-magnetic materials and thus with an assumed permeability of unity. An additional advantage of the transmission line method is there are several methods of decreasing systematic error through a de-embedding process. These de-embedding processes mathematically remove the influence of intermediate transmission lines between the network analyzer faceplate ports and the device under testing (DUT). Moreover, because the DUT is typically a two-part material, powder sample and air, it is possible to de-embed the air from the DUT. This mathematically places the reference planes at the faces of the sample. Additionally, there are several other fundamental advantages stemming from the TEM mode inherent to the design of the coaxial transmission line that decreases the systematic error. Ultimately, the coaxial transmission line method provides a consistent test setup that allows users to remove the majority of the systematic error in their particular system. This provides a basis for comparing values across the literature that have been generated by other research laboratories.

A survey of the current literature reveals several material preparation methods used in the measurement of powdery material in a coaxial transmission line [6]-[9]. These methods include sintering pellets of powdery materials, die pressing powders into pellets, and casting powders into paraffin. Each of these methods has its own drawbacks, resulting in the inconsistency of reported values. Air inclusion is one of the major challenges in sample preparation; no matter how well ordered the powder, there will be air present due to the packing limitation of the powders. This inevitably leads to variability during testing [6]. The incorporation of air becomes more of an issue when there is a wide distribution of particle sizes that result from the incoherent variability of packing fraction. Other scenarios where there is a heterogeneous mixture of particle sizes further complicate the analysis by introducing density variability [7]–[9]. Localized regions of higher density will affect the measured material properties and reflection of the measured scattering parameters.

Sintering is a process where the powders are temporarily held in place and then heated in an oven to cause partial melting and re-solidification of neighboring particles. Because of a change in the local environment of the particles, the dielectric properties of sintered particles differ from their pre-sintered form. Sandi et al. demonstrated that for barium titanate ($BaTiO_3$) that the dielectric properties increase with increasing sintering temperature [10]. While sintering can form samples with good mechanical properties, often the microstructure and subsequently the dielectric properties are influenced. [11], [12]. Costa et al. confirmed the change in particle morphology post-sintering using ultrasound [13]. In addition to changes in local density and morphology of the particles, sintering typically increases the intergranular electrical continuity. Changes in the continuity lead to larger currents and are manifested in changes in both the real and imaginary part of the dielectric constant.

The approach selected for this study, that was aimed at minimizing a change in the intergranular continuity of the material, is to cast the powder particles into a paraffin matrix. In addition to retaining the continuity of the as-received powder, this method is often less capital intensive and faster than sintering. While the mechanical properties are not as good as for sintered particles this is often advantageous for the coaxial test fixture to ensure good contact between electrodes.

By introducing a second phase, paraffin in this case, into the mixture it is necessary to derive the particles' dielectric properties from the two-component mixture. There are several equations available in the literature that have been developed to determine the dielectric constant from a two-part mixture. [14]–[16] The remainder of this study will take a closer look at these equations by conducting a series of controlled experiments at known volume fractions. These results will also be compared with modeling results to gain a better understanding of the theoretical interactions of the different phases and to provide a comparison between analytical equations and modeling results.

## II. Experimental Method

This study will focus on four materials at four-volume loadings. The frequency of interested is between 1 and 10GHz.

### A. Materials and Sample Preparation

The powdery materials selected for this study included silicon dioxide ($SiO_2$ Sigma-Aldrich, 99.9% purity), aluminum oxide ($Al_2O_3$, Inframat Advanced Materials, 99.7%), cerium (IV) oxide ($CeO_2$, Alfa Aesar, 99.5%), and anatase titanium (IV) oxide ($TiO_2$, Acros Organics, 99.5). These materials were selected based on their well-documented dielectric constants at room temperature (RT) and their non-magnetic attributes. The reported values of their relative dielectric constants are 3.5, 8.5, 21.3, and 70. [17], [18] These materials were selected because they encompass a wide range of dielectric constants that are representative of known catalytic powders. Figure 1 is an experimental confirmation of their dielectric properties and their frequency-independence within the frequency range of interest.

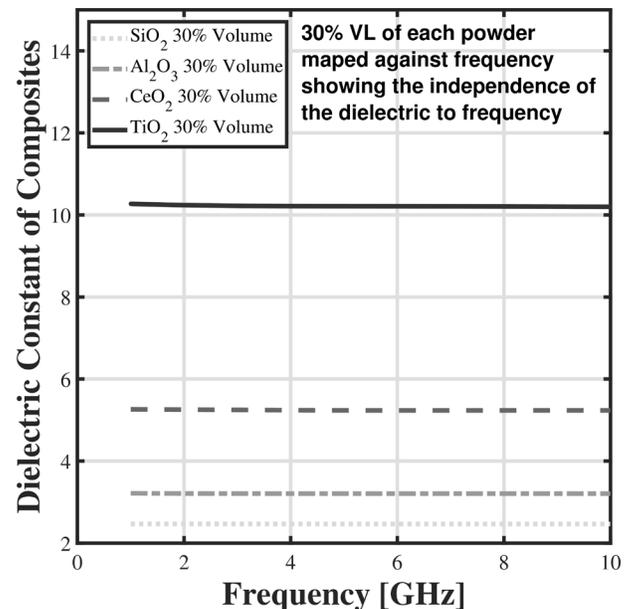

Figure 1. Volume loadings of 30% composites as a function of frequency. All materials investigated in this study were frequency independent over the range of 1-10GHz at every volume loading.

The powder particles were combined with a paraffin matrix at various volume loadings. Volume loading is defined as the volume percentage of powder within the mixture. Paraffin was selected as the matrix material because it is frequency independent at the frequencies of interest with a relative dielectric constant of approximately 2.3. The powder volume loading was selected to be 5, 10, 20, and 30 percent. Volume percentages of the paraffin and powder were calculated prior to combination. Powders were combined with solid paraffin pellets in a beaker and heated to 70°C using a water bath. Once the paraffin was fully melted, the mixture was thoroughly mixed using mechanical agitation and cast into a 3D printed inverted test cell. A photograph of the inverted test cell without the outer sleeve is shown in Figure 2A. The inverted test cell was designed such that the cast plug would make good contact with both the inner and outer electrode. The composite was cooled to room temperature using ambient air and cut to the desired length of 1.0 cm. This length was selected such that it was not equal to the half wavelength of the frequency of interest [19], which was 3.2 cm in this study.

To confirm that the powdery material was not settling to the bottom of the mold during the solidification process, a micro-CT scan of the sample was taken. Figure 2D is the reconstructed image of an alumina/paraffin sample. The dark green color is associated with a higher density alumina phase. It was visually confirmed that there was no substantial settling of the alumina phase. Moreover, it was further confirmed that there were not any noticeable voids or pores that could lead to a ternary phase in the mixture.

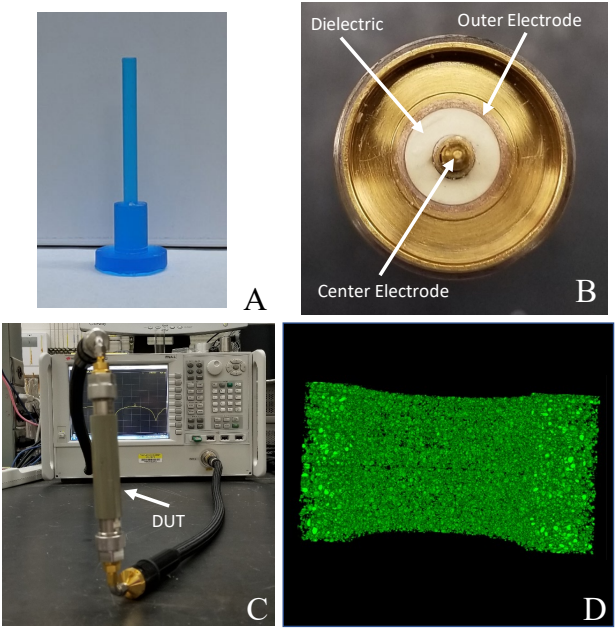

Figure 2. (A) A 3D printed inverted mold used to form sample. (B) Sample loaded into the transmission line. (C) Coaxial airline connected with the VNA. (D) A 3D reconstruction of an $Al_2O_3$ powder and paraffin sample taken with a micro-CT scanner.

### B. Dielectric Testing

All measurements reported in the study were made using a 0.70 cm diameter coaxial airline (HP model no. 85051-60010) connected to a Keysight N5231A PNA-L microwave network analyzer known as a VNA, shown in Figure 2C. Prior to testing the manufacturer's suggested method of calibrating the transmission was followed. This includes a de-embedding process to minimize the systematic error associated DUT.

To take a measurement the sample was loaded into the airline with the center electrode in place, as shown in Figure 2B. All interfaces between the airline and cable were thoroughly cleaned using isopropyl alcohol and dried using dry compressed air. Information about the sample length was provided to the VNA software and the scattering parameters (S-parameters) were measured at 1601 equally spaced points within a frequency range of 1-10GHz. The relative dielectric constant for each sample was calculated from the measured scattering parameters using the Nicholson-Ross-Weir polynomial method, which is built into the VNA Keysight Software [17]. This returns a relative dielectric constant as a function of frequency for the composite mixture of powder and paraffin.

### III. ANALYTICAL MIXING EQUATIONS

This study investigated ten different mixing equations across four volume loadings for four different powder materials. All ten of the equations investigated are provided in Equations 1-10. The existence of a large number of equations stem from the difficulty in creating a single mixing equation for a two-phase homogenous composite as no exact solution exists [1]. The approach taken in this study was to use powders with known dielectric constant and measure precise volume fractions to experimentally determine the fitness of each of the mixing equation.

There are several factors that contribute to the variability of experimentally determined dielectrics. Dielectric constants for the same material will generally fall within a range of values that have a well-defined standard deviation. The Hashin-Shtrikman bounds are generally considered the limits for determining the effective permittivity of a mixture for a given volume fraction [1]. There are also wider bounds known as the Wiener bounds which represent the absolute bounds of a mixture based on the material orientation [1]. This random error is a result of several physical attributes of the powders. These include size, shape, and intergranular effects from neighboring individual grains of powder. These intergranular effects are a lesser understood problem of powdery materials but are an area of ongoing research. [21]

The most well-known empirical equations for predicting the dielectric constant of a two-phase homogenous compound are provided in Equations 1-10. It is important to understand that each of these equations were empirically derived for a certain frequency range and dielectric constant range with respect to the mixture. Often one equation is better suited to predict a certain range of dielectric properties and frequencies. The original application of each of these equations can be found in their associated manuscripts where they are derived.

$$\varepsilon_{mix} = V_p \varepsilon_p + V_m \varepsilon_m \quad (1)$$

$$(\varepsilon_{mix})^{-1} = V_p (\varepsilon_p)^{-1} + V_m (\varepsilon_m)^{-1} \quad (2)$$

$$\varepsilon_{mix} = \varepsilon_p^{V_p} + \varepsilon_m^{V_m} \quad (3)$$

$$\varepsilon_{mix} = \varepsilon_p^{V_p} \varepsilon_m^{V_m} \quad (4)$$

$$\varepsilon_{mix}^{1/3} = V_m \varepsilon_m^{1/3} + V_p \varepsilon_p^{1/3} \quad (5)$$

$$\varepsilon_{mix}^{1/2} = V_m \varepsilon_m^{1/2} + V_p \varepsilon_p^{1/2} \quad (6)$$

$$\varepsilon_{mix} = \varepsilon_m \left[1 + \frac{V_p(\frac{\varepsilon_p}{\varepsilon_m} - 1)}{V_p + \left(\frac{V_m}{3}\right)\left[\frac{\varepsilon_p}{\varepsilon_m} V_m + V_p + 2\right]}\right] \quad (7)$$

$$\varepsilon_{mix} = \varepsilon_m \left[1 + \frac{V_p(\varepsilon_p - \varepsilon_m)}{\varepsilon_m + nV_p(\varepsilon_p - \varepsilon_m)}\right] \quad (8)$$

$$\varepsilon_{mix} = \varepsilon_m + 3 * V_p \varepsilon_m \left(\frac{\varepsilon_p - \varepsilon_m}{\varepsilon_p + 2\varepsilon_m - V_p(\varepsilon_p - \varepsilon_m)}\right) \quad (9)$$

$$\varepsilon_{mix} = \frac{\varepsilon_m V_m + \varepsilon_p V_p \left[\frac{3\varepsilon_m}{(\varepsilon_p + 2\varepsilon_m)}\right]\left[1 + \frac{3V_p(\varepsilon_p - \varepsilon_m)}{(\varepsilon_p + 2\varepsilon_m)}\right]}{V_m + \frac{V_p(3\varepsilon_m)}{(\varepsilon_p + 2\varepsilon_m)}\left[1 + \frac{3V_p(\varepsilon_p - \varepsilon_m)}{(\varepsilon_p + 2\varepsilon_m)}\right]} \quad (10)$$

Equation 1-10 are known as the following, Parallel (Eq. 1), Series (Eq. 2), Logarithmic (Eq. 3), Lichtenecker (Eq. 4), Looyenga (Eq. 5), Birchak (Eq. 6), Poon-Shin (Eq. 7), Electro-Magnetic Theory (EMT) (Eq. 8), Maxwell-Garnet (Eq. 9), and Jayasundere-Smith (Eq. 10) mixing rules respectively [1], [16], [22]–[24]. Where $\varepsilon_{mix}$, $\varepsilon_m$, and $\varepsilon_p$ are the dielectric constants of the mixture composite, matrix material (paraffin), and particle material (powder). $V_p$ is the volume loading of the particle and $V_m$ is the volume loading of the paraffin. The EMT mixing rule uses a particle shape factor denoted by *n*. The assumed *n* value used for the entirety of this study is 1.05. This value was determined by the particle shape using the methodology laid out by Sihvola [1]. Every equation except for the EMT mixing rule is dependent on only four variables.

For each of the four powders and four volume loadings investigated the particle's dielectric properties were determined using all ten equations. A quantitative and qualitative assessment of the fitness of each equation was carried out. A numerical analysis of percent error between the predicted dielectric constant derived from the mixture and the known value for the given powder was considered.

## IV. RESULTS AND DISCUSSION

A plot of the dielectric constants for the composite mixtures versus volume loading for each of the materials is provided in Figures 3-7. Each of these plots includes 1601 experimentally measured data points between 1-10 GHz that are provided from the VNA. These measurements are represented as a single data point with error bars at each of the four volume loadings. The error bars are determined from the experimental measurement of 6 samples at each volume loading. The confidence interval was determined based on a t-distribution with 95% confidence. The other lines in each of the plots are associated with individual mixing equations.

### A. Dielectric Measurement of Powdery SiO2

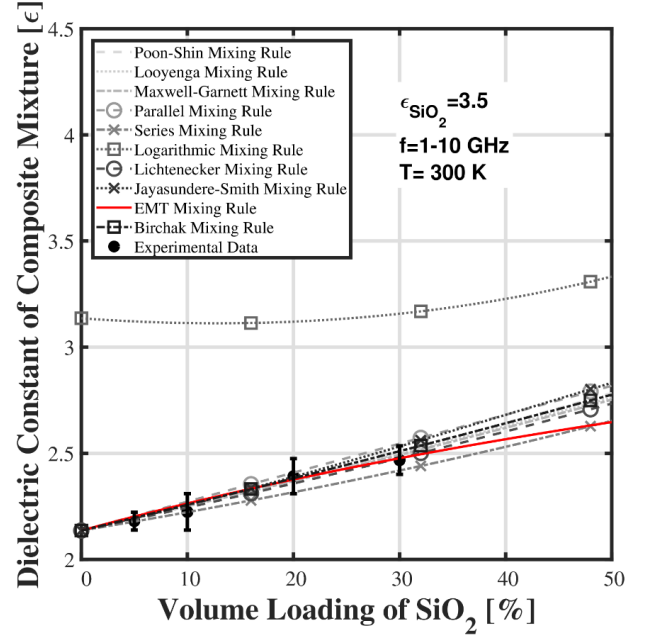

Figure 3. Dielectric constant of SiO$_2$ powder in paraffin matrix. Theoretical dielectric value of 3.5 was used for mixing equations. EMT Mixing Rule was best fit to experimental data with n=1.05.

The dielectric constant of silica (SiO$_2$) is just slightly greater than that of paraffin resulting in a dielectric inclusion ratio, $\varepsilon_p/\varepsilon_m = 1.5$, $\varepsilon_m = 2.3$. Figure 3 is a scatterplot of the dielectric constant of the powder/paraffin composite versus the volume loading mapped against the predicted dielectric properties from all 10 equations. The experimental data depicted as solid dots in Figure 3 is based on the averaging of six samples at four different volume loadings. The error bars associated with each of these points is based on a calculated standard deviation.

As illustrated in Figure 3, all the mixtures except the Logarithmic mixing rule followed the overall trend of the experimental data. All other equations are within 5% mean absolute percent error (MAPE) for predicting the average dielectric constant at each volume loading. At lower volume loadings (<20%) the error is less than 2% for all equations except Logarithmic. The percent error across all volume loading for the best fit equation of EMT is shown in Table 1 as 0.94%. At small $\varepsilon_p/\varepsilon_m$ the mixture equations behave mathematically the same with a linear dependence of volume fraction. At volume loadings above 30%, the spread between equations is greater than the error bars. This provides justification for maintaining the volume loading below 30% for small inclusion ratios.

It is important to point out that as the volume loading approaches 100% the value should converge on the particles dielectric value, in the case of SiO$_2$ this value is 3.5. Looking at Equation 3, which represents the Logarithmic equation, some claims can be made on why this equation resulted in a large error for small $\varepsilon_p/\varepsilon_m$. The reason being that when the exponent terms in the expression are similar in magnitude the bases of the exponent are effectively additive. This is apparent in Figure 3 because across the volume loading range the Logarithmic equation is nearly equal to the pure powder values. It will be demonstrated for higher $\varepsilon_p/\varepsilon_m$ that this is less apparent because one of the terms dominates the expression giving rise to a power-law relationship versus volume loading.

### B. Dielectric Measurement of Powdery $Al_2O_3$

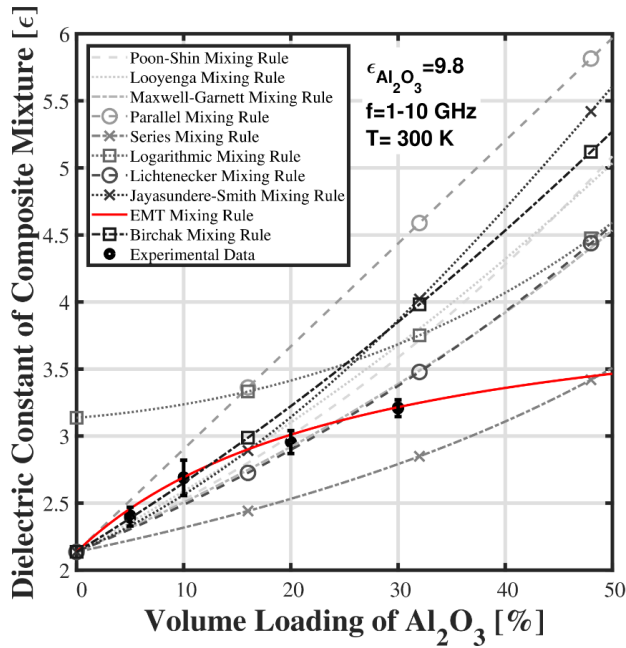

Figure 4. Dielectric constant of $Al_2O_3$ powder in paraffin matrix. Theoretical dielectric value of 9.8 was used for mixing equations. EMT Mixing Rule was best fit to experimental data.

Figure 4 is a plot of the dielectric constant versus volume for an alumina ($Al_2O_3$) powder mixed with paraffin matrix. Alumina has a known dielectric cost of 9.8 resulting in an increased dielectric inclusion ratio of $\varepsilon_p/\varepsilon_m$=4.3. As seen in Figure 4 in comparison to Figure 3, there is a greater spread in the predicted values of the mixing equations. The black dots in Figure 4 are based on six experimentally measured values. As seen in Figure 4, at lower volume loadings, the majority of mixture equations are accurate with the exception of the Logarithmic and Series Mixing equations. However, at higher volume loadings, only one of the equations follows the trends of the experimental data. This equation is the EMT Mixing Rule, shown as a solid line red line Figure 4. The MAPE of the EMT is shown in Table 1 as 1.16% across all volume loadings.

The deficiency in the Logarithmic equation can be associated with the previous mention comments made in the $SiO_2$ discussion. For small $\varepsilon_p/\varepsilon_m$, Equation 3 suffers from a large shift in the y-axis as a result of one of the terms in the equation dominating the summation. The other equation that failed to follow the trends in Figure 4 was the Series Mixing Rule. Looking at Equation 2, which represents the Series Mixing equation, it is representative of the summation of two inverse terms. This means the output will be limited by the smallest inverse component. In the case of low volume loading, this would be the paraffin and it is not until higher volume loadings that $Al_2O_3$ becomes the dominating component. This is visually represented in Figure 4 as an increase in the Series Mixing Equation with increasing volume loading.

In the case of the EMT Mixing Rule, which performed best at matching the overall trends, there is concern over higher volume loadings above 40%. At these higher volume loadings, the trend underpredicts the value. The shift in the trend above 40% volume loads can be directly correlated to the shape factor term associated with Equation 8. This provides justification for maintaining the volume loading below 30%.

For the associated inclusion ration for $Al_2O_3$ and paraffin, it is best practice is to keep the volume loading below 30% and to use the EMT mixing rule with the empirical selection of a n-value or to use one of the other mixing rules except Series Mixing or Parallel Mixing.

### C. Dielectric Measurement of Powdery $CeO_2$

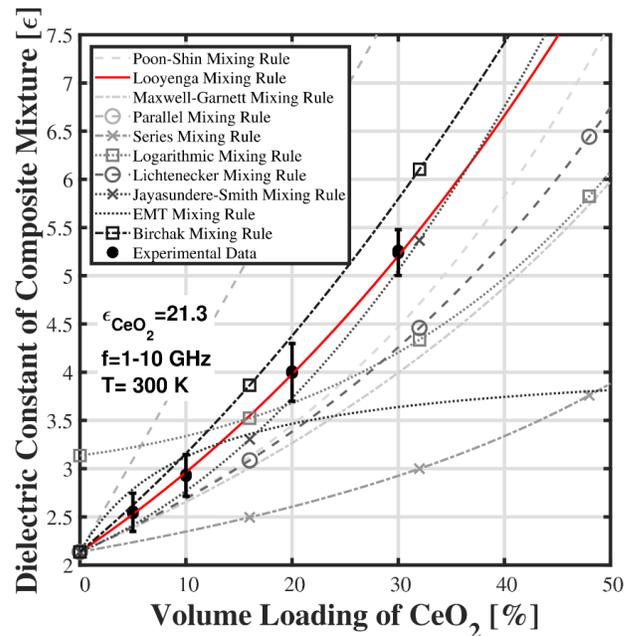

Figure 5. Dielectric constant of CeO2 powder in paraffin matrix. Theoretical dielectric value of 9.8 was used for mixing equations. EMT Mixing Rule was best fit to experimental data.

Figure 5 is a plot for cerium oxide ($CeO_2$) with a dielectric constant of 23. The associated dielectric inclusion ratio for cerium oxide mixed with paraffin is $\varepsilon_p/\varepsilon_m = 10$. The trends of the equations do not significantly deviate from those seen for $Al_2O_3$, with the trend displaying a slight positive curvature with increased volume loading. It is noted that the Logarithmic equation is now doing better at correlating with the experimental values as the inclusion ratio is increased from near unity. Comparing the experimental values depicted as black dots in Figure 5 to each of the equations, the Looyenga Mixing Rule does the best at minimizing the error. Across all volume loadings the Looyenga mixing equation results in a 0.76% MAPE as shown in Table 1. This is a change in the best-fit equation from the previous powders. It is reasoned that the dielectric inclusion ratio is a good indicator for selection of the equation. Inspection of the equation for Looyenga mixing rule, Equation 5, finds that the equation is the summation of two power laws. Where the exponent is fixed at 1/3. If Equation 5 is rearranged as follows,

$$\varepsilon_{mix}^{1/3} = [V_m + V_p(\varepsilon_p/\varepsilon_m)^{1/3}]\varepsilon_m, \quad (11)$$

it becomes apparent that the expression is heavily dependent on the ratio of dielectric constants between the particle and matrix. The particle will not significantly contribute to the trend until the ratio is above 10, which is realized in the case of $CeO_2$ but not the previous powders.

### D. Dielectric Measurement of Powdery $TiO_2$

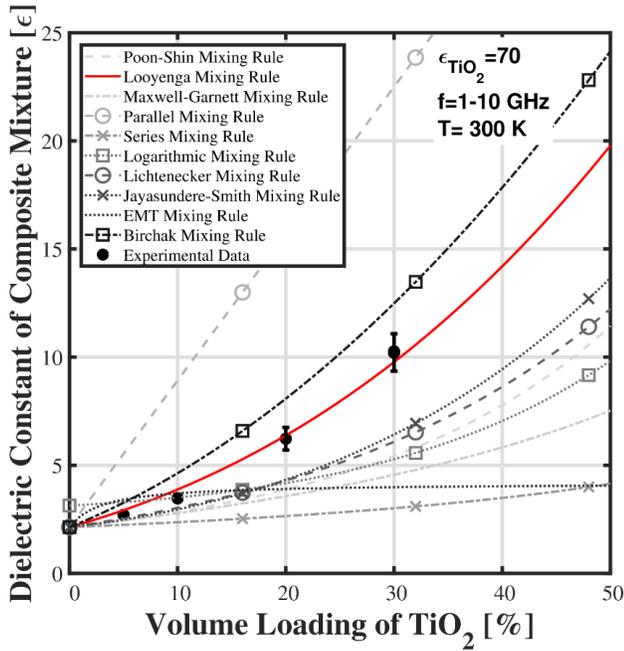

Figure 6. Dielectric constant of $SiO_2$ powder in paraffin matrix. Theoretical dielectric value of 3.5 was used for mixing equations. EMT Mixing Rule was best fit to experimental data.

Figure 6 is a plot for titanium dioxide ($TiO_2$) with a dielectric constant of 70 and an inclusion ratio $\varepsilon_p/\varepsilon_m = 30$. It is qualitatively noted from Figure 6 that the trends of the lines are becoming more spread-out suggesting that it is of increased importance to select the correct equation at these higher inclusion ratios. In addition to larger dispersion amongst all the equations, the curvature is also increasing as a function of volume loading. This increase in curvature is attributed to the more significant contribution of the particles' dielectric to the overall mixture's dielectric constant predicted by the mixing equations. It is seen from Figure 6 that the best-fit equation is the Looyenga Mixing Rule. This is the same finding as to the cerium oxide case reaffirming that at inclusion ration above 10 Looyenga Mixing Rule is the best-fit equation.

As seen in Figure 6, at volume loadings at and above 20% only the Looyenga mixing rule can accurately predict the compound dielectric constant. When the volume loading is 10% the EMT, Logarithmic, and Looyenga mixture equations pass near the experimental point. However, the EMT and Logarithmic equation do not map the trend of the dielectric constant above 10%.

Unlike the previous powders investigated, when the volume loading of $TiO_2$ is at 5 and 10 percent, none of the equations besides Looyenga can predict the dielectric constant at a less than 5% error. Moreover, only the Looyenga Mixing Rule can get within 20% error for all of the powder mixtures investigated in this study. Table 1 summarizes the results for all volume loadings investigated in this study. Therefore, if the inclusion ratio is unknown it is best practice to form a mixture with 10-20% volume fraction and to use the Looyenga Mixing Rule.

## V. MODELING RESULTS

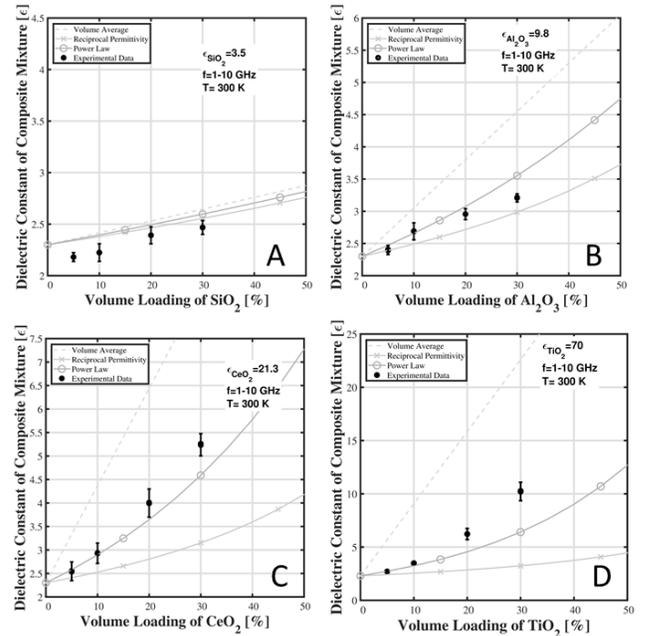

Figure 7. EM FEM simulations for the dielectric constants of (A) $SiO_2$, (B) $Al_2O_3$, (C) $CeO_2$, and (D) $TiO_2$ powders in a paraffin matrix as a function of volume loading. Theoretical dielectric constant values of 3.5, 9.8, 21.3 and 70 were used respectively for mixing the constitutive material while 2.3 was used for the paraffin matrix. The simulation results for all two-phase composites are mapped against experimentally collected data.

Figure 7 illustrates the Finite Element Method (FEM) COMSOL results for the four powders investigated in this study. COMSOL has three different option for the averaging of the dielectric constant of multi-phased materials, volume averaging, reciprocal permittivity averaging and power law averaging. A comparison of these three EN results with the analytical equations presented in this paper show that the modeling software is utilizing three of the more simplistic equations. The volume averaging, reciprocal permittivity averaging, and power law results correspond to the results of Equations 1, 2 and 4.

All four modeling results show the same increasing trend of dielectric constant as volume loading increases as those equations for the given powder. It can be seen from Figure 7 that the power law volume mixing is only accurate for low volume loadings (<20%) and for those powders with a dielectric inclusion ratio above 1.5. The MAPE for low volume loadings of $SiO_2$, $Al_2O_3$, $CeO_2$ and $TiO_2$ when using the modeled results are 7.67, 2.09, 1.26 and 3.86 respectively. While these results are not as accurate as the EMT or Looyenga mixture equations as discussed above. The available mixing models in the EM FEM software capture the general trend of increasing dielectric with increasing volume fraction addition. However, the lack of accuracy does draw to concern if these relationships are used to solve the inverse of problem. In the case of solving the inverse it is best to use the EMT mixing rule or Looyenga mixing equation as outlined in Table 1 and Table 2.

## VI. Method Guidelines

In providing a systematic approach to measuring unknown powders using a transmission line method the following list of steps has been enumerated for guidance. Table 2 also shows a simplified guideline for which mixture equation to utilize in determining the powders dielectric property from that of the composite mixture.

The following procedure is suggested when measuring the dielectrics of powdery materials:

1) Mix powder particles and matrix material together at 10% volume loading. The suggested matrix material is paraffin.
2) Approximate the inclusion ratio of your mixture ($\varepsilon_p/\varepsilon_m$).
3) Calibrate the VNA and de-embed air from the transmission line.
4) Measure the composite dielectric properties using a coaxial transmission line.
5) Select appropriate equation: EMT if $\frac{\varepsilon_p}{\varepsilon_m} < 10$ and Looyanga if $\frac{\varepsilon_p}{\varepsilon_m} \geq 10$. If the inclusion ratio is unknown, select Looyanga.
6) Using the selected mixing equation based on the answer to question 5, calculate powder's dielectric properties.

## VII. Conclusion

This study investigated a variety of powder materials with a wide range of dielectric properties in a coaxial transmission line. The objective was to gain an understanding of when to best utilize the mixture equations in order to calculate a powder's dielectric constant ($\varepsilon_p$) based on the experimental measurement of a composite mixture. Each of the powders were mixed with a paraffin matrix material for easy control. The dielectric properties for these pure powders were known prior to measuring the mixture's effective properties. Allowing the mixture equations to be used in a manner that they could predict the effective dielectric properties. However, in practice, these equations would be used in reverse to predict the unknown dielectric of the powder. It was found that using a traditional EM modeling technique was not a viable practice for predicting the dielectric property of any two-phase composite. Instead, it was found that the inclusion ratio, $\varepsilon_p/\varepsilon_m$ proved to be a good metric for selecting the appropriate mixing equation that minimizes the error. Table 1 and Table 2 provide a summary of the findings from this study. For inclusion ratios below 10, the sensitivity of the powders' dielectric to the measured effective dielectric on the mixture was less and therefore required a rigorous mixing equation. This results in the selection of the EMT Mixing Rule for inclusion ratios below 10. For inclusion ratio above 10, the sensitivity with volume fraction changed and the best-fit mixing equation was found to be the Looyenga Mixing Rule.

In addition to selecting the correct mixing equation, this study also found that care must be taken in selecting the correct volume loading or volume fraction of powder. It is recommended that the volume loading is below 20%. It was found that the error at 20% volume fraction was 10% or less for all the powders considered.

When selecting powders, smaller particle sizes (<65 μm) were preferred for composite preparations. The particle size of 30 μm had the best distribution within the paraffin matrix as well as typically having the most accurate volume loadings. The smaller particle sizes helped to avoid any particle clustering that would have interfered with the dielectric constant of the mixture.

| Powder | Suggested Mixing Equation | Percent Error (%) |
|---|---|---|
| $SiO_2$ | EMT (n=1.05) | 0.94 |
| $Al_2O_3$ | EMT (n=1.05) | 1.16 |
| $CeO_2$ | Looyenga | 0.76 |
| $TiO_2$ | Looyenga | 6.21 |

Table 1: The best mixing equation for each powdery material studied and the associated averaged error over the volume fractions studied between the equation and the experimental values. The EMT mixing equation is recommended over the Maxwell-Garnett for simplicity despite a statistically insignificant improvement of error when using the latter equation.

| Ratio of Dielectric Properties, $\varepsilon_p/\varepsilon_m$ | Suggested Mixing Equation |
|---|---|
| 1 - 10 | EMT (n=1.05) |
| 10-100 | Looyenga |

Table 2: Suggested mixing equation based on the ratio of powder's dielectric properties to matrix material's dielectric properties.